\documentclass[10pt]{article}

\usepackage{fullpage}
\usepackage{setspace}
\usepackage{parskip}
\usepackage{titlesec}
\usepackage{xcolor}
\usepackage{lineno}

\PassOptionsToPackage{hyphens}{url}
\usepackage[colorlinks = true,
            linkcolor = blue,
            urlcolor  = blue,
            citecolor = blue,
            anchorcolor = blue]{hyperref}
\usepackage{etoolbox}
\makeatletter
\patchcmd\@combinedblfloats{\box\@outputbox}{\unvbox\@outputbox}{}{%
  \errmessage{\noexpand\@combinedblfloats could not be patched}%
}%
\makeatother

\usepackage{natbib}

\renewenvironment{abstract}
  {{\bfseries\noindent{\abstractname}\par\nobreak}\footnotesize}
  {\bigskip}

\titlespacing{\section}{0pt}{*3}{*1}
\titlespacing{\subsection}{0pt}{*2}{*0.5}
\titlespacing{\subsubsection}{0pt}{*1.5}{0pt}

\usepackage{authblk}

\usepackage{graphicx}
\usepackage[space]{grffile}
\usepackage{latexsym}
\usepackage{textcomp}
\usepackage{longtable}
\usepackage{tabulary}
\usepackage{booktabs,array,multirow}
\usepackage{amsfonts,amsmath,amssymb}
\providecommand\citet{\cite}
\providecommand\citep{\cite}

\newif\iflatexml\latexmlfalse

\AtBeginDocument{\DeclareGraphicsExtensions{.pdf,.PDF,.eps,.EPS,.png,.PNG,.tif,.TIF,.jpg,.JPG,.jpeg,.JPEG}}

\usepackage[utf8]{inputenc}
\usepackage[english]{babel}

\begin{document}

\title{A Blockchain Based Liability Attribution \\  Framework for Autonomous Vehicles.}

\author[1]{Chuka Oham}%
\affil[1]{University of New South Wales, Sydney, Australia}%

\author[1]{Salil S. Kanhere}%

\author[2]{Raja Jurdak}%
\affil[2]{Data 61, CSIRO, Australia}%

\author[1]{Sanjay Jha}%

\begingroup
\let\center\flushleft
\let\endcenter\endflushleft
\maketitle
\endgroup

\selectlanguage{english}
\begin{abstract}
The advent of autonomous vehicles is envisaged to disrupt the auto insurance liability model. Unlike the current model, wherein in the event of an accident, the liability is largely attributed to the driver, autonomous vehicles necessitate the consideration of other entities in the automotive ecosystem such as the auto manufacturer, software provider, service technician and the vehicle owner for liability attribution. The proliferation of sensors and connecting technologies in autonomous vehicles enables an autonomous vehicle to gather sufficient data for liability attribution, yet increased connectivity exposes the vehicle to attacks from interacting entities such as an automaker with significant understanding of the inner workings of the autonomous vehicle or the vehicle owner with unrestricted access to the internal network of the vehicle. These possibilities motivate potential liable entities to repudiate their involvement in a collision event to evade liability. While the data collected from vehicular sensors and vehicular communications is an integral part of the evidence for arbitrating liability in the event of an accident, there is also a need to record all interactions between the aforementioned entities to identify potential instances of negligence that may have played a role in the accident. Furthermore, autonomous vehicle sensors are envisaged to collect significant amount of data including personal identifiable information about a vehicle owner which threatens his privacy. In this paper, we propose a BlockChain (BC) based framework that integrates the concerned entities in the liability model and provides untampered evidence for liability attribution and adjuducation. We first describe the liability attribution model,  identify key requirements for our proposed BC-based framework and describe the adversarial capabilities of  entities. Also, we present a detailed description of relevant data contributing to evidence and further describe how they are stored and used as evidence. Our framework uses permissioned BC, to restrict access to relevant entities and partitions the BC to further tailor data access to relevant BC participants. Finally, we conduct a security analysis to verify that the identified requirements are met  and resilience of our proposed framework to identified attacks.%
\end{abstract}%

\section*{Introduction and Motivation}
The increased adoption of autonomous vehicles is expected to disrupt the auto insurance industry. Vehicle autonomy shifts some or all driving decisions from the human driver to the vehicle, which will require a complete overhaul of existing liability models. Autonomous vehicles are instrumented with a wide array of sensors, embedded computers and communicating technologies to enable better perception of the environment and facilitate independent decision making to avert road transportation hazards. However, the possibility to make independent driving decisions introduces the challenge of liability attribution. A realistic scenario is described in [1] where an autonomous vehicle driver was killed in an accident. The appropriation of liability in this scenario was a significant challenge because the auto manufacturer claimed it did not receive any log about the incident and could not ascertain the drive mode of the vehicle at the time of the accident while the family of the deceased claimed otherwise. Whereas current liability attribution usually assigns blame to the driver, liability for autonomous vehicles is to be shared between multiple entities responsible for the operation of the vehicle including the auto manufacturer, maintenance service provider, driver and software provider. The possibility of shared liability therefore highlights the urgent need for a new auto insurance liability framework.  \\
The authors in [2] introduce the following two categories of liability attribution for autonomous vehicles: i) \textit{Product liability} which refers to damages due to product defects such as design failure and manufacturing failures. Liability is either attributed to the auto manufacturer when an accident occurs and the vehicle is in autonomous mode or to the software provider when a software program is deemed to have led to the accident by causing erroneous action. A service technician is liable if an accident is traced to its last action on the vehicle. ii) \textit{Negligence liability} which refers to damages due to neglects to execute an action. Liability is attributed to a vehicle owner when he fails to execute an instruction (software update) from an auto manufacturer or software provider. An auto manufacturer is however considered negligent if the accident could have been prevented by a human driver. Therefore, given the associated liability costs, e.g., compensation and potential tarnishing of reputation, an entity that is deemed liable may be strongly motivated to deny its actions thus making liability attribution challenging. Also, an autonomous vehicle houses multiple data gathering sensors such as  Light Detection and Ranging (LIDAR) for generating a 3D map of its surroundings, cameras to read speed limit signs and watch lane marks to prevent drifting, GPS and tire pressure monitoring sensors. Given the high exposure of the autonomous vehicle to external communication, an auto manufacturer with significant understanding of the inner workings of the vehicle could remotely tamper with sensors generating data for evidence to evade liability. Remote exploitation has already been successfully demonstrated for a modern day vehicle [3]. An autonomous vehicle is envisaged to generate significant amount of data, some of which could reveal private information about the driver and passenger [2]. This introduces privacy concerns when data generated for evidence is either retrieved or transmitted to decision makers for dispute settlement.   \\
In addition, gathering data that constitutes evidence for liability attribution is challenging. The National Road and Motorists Association (NRMA) [4], Australia identified data constituting evidence as time of event, date, location, description of event, witness testimonies and pictorial evidence. In the traditional liability model, when an accident occurs, forensic investigators visit the site to obtain physical evidence such as impact of damage, vehicle position and vehicle heading. They also extract data from the Event Data Recorder (EDR); a black-box installed in the vehicle that  stores pre and post collision data such as speed, accelerator angle, usage of safety systems etc. These data enable a vehicle to prove its recent behaviour for liability purposes. However, such data is not sufficient for liability attribution as it cannot be used to prove the interaction between a vehicle and other potentially liable entities to reveal potential instances of negligence which could also be the cause of the accident. Thus, new mechanisms are needed to record relevant data exchanges between liable entities as evidence to ascertain if the accident was due to negligence. \\
Furthermore, given the popular consensus to attribute liability to an auto manufacturer or software provider for product defects and to the vehicle owner for negligence [2], a likely liable entity might be motivated to deny the availability of data or receipt of data contributing to evidence. In the realistic scenario described in [1], the auto manufacturer could not ascertain the cause of an accident because the accident log files were not logged in their server due to severity of damage. The auto manufacturer was also denied access to more information by the vehicle owner to aid its investigation. The consensus to split liability [2] in the autonomous vehicle setting and the vested interest of liable entities in the outcome of liability decisions, highlight the need for a transparent liability framework where such data contributing to evidence is accessible and universally accepted by concerned entities in the liability framework for expediting liability decisions. \\
The emerging Blockchain technology has the potential to underpin a new liability framework for autonomous vehicles as it provides trustless consensus. Blockchain (BC) is an immutable and distributed ledger technology that provides verifiable record of transactions in the form of an interconnected series of data blocks.  Blockchain's immutability is achieved by using the hash of the previous block to chain interconnected data blocks so that changing the content of a single block will require changing the headers of previous blocks up to the genesis block; the first block in the chain. In addition to the immutability and security features of BC, it also offers privacy to nodes in a network and therefore is a useful technology to address aforementioned challenges of liability attribution for autonomous vehicles including the potential alteration of evidence to evade liability, the need for auditability and accessibility of evidence. Nodes in BC are identified using a Public Key (PK) and communication exchanges between nodes are called transactions. The transactions are encrypted using the PK. In BC, every node can verify a transaction in the network by validating the signature on transactions against their PK. This ensures ownership of transactions. A BC node can also change its PK for every transaction for anonymous communication and privacy. \\
There are two kinds of BC: public and permissioned. The distinction between these BC types relates to who is allowed to participate and their capabilities i.e. who maintains the shared ledger and executes the consensus protocol [5].  Public BC, such as BitCoin [5],  enables every participant to create transactions, collaboratively verify them and append validated transactions to the ledger. Permissioned BC such as Hyperledger [6] on the other hand restricts participation by inviting participants to join. Compared to the public BlockChain (BC), in permissioned BC, only designated participants are responsible for maintaining the ledger.  Furthermore, to control user participation, public BC requires significant computational resource to reach consensus on the next block to be added to the BC. In contrast, in a permissioned BC since the identities of all entities are known, there is no need to execute a rigorous consensus protocol, thus reducing the computational overheads. In addition, while nodes in a Public BC are privy to all communication in the network, permissioned BC restricts communication to ensure that only parties to a transaction have knowledge of the transaction [5].\\
Several researchers have recently proposed to use BC in the vehicular context for different purposes. Blackchain [7], a message and revocation accountability system, is a BC based revocation mechanism for secure vehicular communication.  It prevents issuing communication credentials to malicious vehicles and also prevents credential issuing authorities from misbehaving. The authors in [8] presents a BC based secure architecture for automotive use cases such as wireless software updates and dynamic vehicle insurance which allows an insurance company to provide specific services to drivers by monitoring their driving patterns. \\
The main contribution of this work is to propose a new auto insurance liability attribution model for autonomous vehicles and present an efficient data gathering, privacy preserving and secure BC framework for liability attribution for autonomous vehicles. The liability attribution model integrates all concerned entities for relevant data exchanges to facilitate liability decision making including the autonomous vehicle, service technicians, insurance companies, auto manufacturers and the government authorities who make available evidence and makes liability decisions such as the road transport authority and legal authorities (police, courts). Our framework is based on permissioned BC because: 1) of  its access control mechanism for restricting access to data in the ledger; 2) it is scalable as its consensus process does not require expending significant computational resources; and 3) it offers privacy as only authorized entities have access to an evidence.\\
Our proposed framework also facilitates the collection of comprehensive evidence for making liability decisions. Comprehensive evidence in our framework are referred to as transactions and include relevant interactions between an autonomous vehicle and other potential liable entities such as the instruction to execute an update and the execution of the instruction to identify a potential case of negligence; the report of a maintenance officer on the maintenance conducted in an autonomous vehicle; the data generated by an autonomous vehicle in the event of a traffic accident; and data generated by an autonomous vehicle events such as hard braking, over speeding and crossing a red light are triggered. These data are stored in the BC and used for making liability decisions. \\

\section*{Background}
We now provide the necessary background for designing a new liability attribution model for autonomous vehicles. We begin by describing the roles of interacting entities in the proposed model and identify key requirements to guide the design of our proposed BC based framework. \\
Interacting entities in our liability attribution model includes: 
\begin{itemize}
\item \textbf{Autonomous Vehicle (\textit{AV})}: Provides primary evidence needed for forensic analysis to make liability decisions. It is assumed to have a tamper resistant storage for storing data generated by sensors about an accident such as the precise location, speed, time of event, video and picture data of an accident including data received from witnesses ($W_{i}$) via vehicle-to-vehicle (v2v) communication. Details about the data stored as evidence by an \textit{AV} are described in the next section. 
\item \textbf{Auto manufacturer (\textit{AM})}: Periodically receives sensor data information from an \textit{AV} for diagnostics and  scheduling maintenance.  The task of providing updates to vehicles is also subsumed within the \textit{AM}.  It provides over-the-air (OTA) software updates to an \textit{AV} to upgrade its functionality or for bug fixes in an embedded software installed in a data generating sensor. Such instruction to an \textit{AV} is used to identify potential instances of negligence i.e. to check if the \textit{AV} executed the instruction. 
\item \textbf{Witnesses \textbf{($W_{i}$)}}: We refer to other vehicles involved in an accident or vehicles within camera range of the accident as witnesses. Witnesses involved in the accident could also be liable for the accident as such they provide evidence to their insurance companies, auto manufacturers and legal authorities for making liability decisions. Other witnesses within the camera range could transmit their recorded perception to a road side infrastructure via vehicle-to-infrastructure (v2i) communication to apprise other vehicles about the event. The road side infrastructure could also transmit the received accident data to the government transport authority which passes on the data to law enforcement agencies as evidence for dispute settlement in court. 
\item \textbf{Service Technician \textit{(ST)}: } Provides maintenance services for the autonomous vehicle and provides a report as a proof of service to the vehicle.
\item \textbf{Insurance Company (\textit{IC})}: Receives evidence from its client (\textit{AV}), legal authorities and the governmental transport authorities and processes the received evidence to make liability decisions. It also pays compensation for damages and injuries on behalf of client if responsible for an accident. 
\item \textbf{Government Transport Authority (\textit{GTA})}: We assume that a \textit{GTA} would expand their modern role as license issuing authority and operators of roadside infrastructures such as stop signs and traffic lights by issuing smart roadside infrastructures and cryptographic credentials needed to secure v2v and v2i communication.  Secure vehicular communication allows an autonomous vehicle receive data from only vehicles with valid communication credentials. 
\item \textbf{Legal Authority (\textit{LA})}: Legal Authorities comprise of the law enforcement agencies (police and court). They receive and analyze evidence from likely liable entities  for dispute settlement. They also provide evidence to insurance companies to facilitate payment payment of compensations. 
\end{itemize} 
We refer to \textit{GTA} and \textit{LA} as Dispute settlement Authorities (\textit{DA}) as they cooperate to identify a liable autonomous vehicle for dispute settlement [9]. This is to ensure that the identity of an autonomous vehicle stays hidden and can only be revealed when \textit{GTA} and \textit{LA} cooperate to prevent data abuse. Figure 1 describes relevant interactions between entities in the proposed liability attribution model. 
\begin{figure}[h]
\centering
\includegraphics[width=0.70\textwidth]{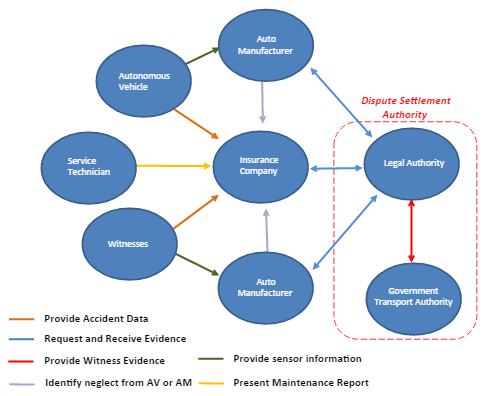}
\caption{Proposed Liability Attribution Model.}
\end{figure}

\textbf{Key Requirements} \\
We identify specific requirements to address the aforementioned challenges for our proposed framework. These requirements will guide the design of our blockchain based liability attribution framework for autonomous vehicles. 
\begin{itemize}
\item \textbf{Evidence Integrity:} Entities involved in the liability model may attempt to alter previously submitted evidence to deny responsibility thus the liability model must ensure that previously submitted evidence remains unchanged. 
\item \textbf{Secure storage:} Entity interactions contributing to evidence including a description of historical behaviour of the vehicle should be available for making liability decisions.
\item \textbf{Non-repudiation:} By providing proof of data origin and integrity, participants would not be able to deny their actions.
\item \textbf{Decentralization: } Given multiple sources of liability in the liability attribution model, no single entity should be the ultimate source of truth on evidence for making liability decisions. Evidence must be verified and accepted as valid by all verifiers. 
\item \textbf{Authorization:} Data exchanges between entities should be based on relevance to mutual communicating entities and the need to know. 
\item \textbf{Privacy:} Autonomous vehicles may collect and maintain identifying information about the owner or passenger of the vehicle for a variety of purposes, such as to authenticate authorized use, or to customize comfort, safety, and entertainment settings. This information likely will be able to identify owners and passengers and their activities with a high degree of certainty. It is thus important that such personal information is protected. However, it should be possible to identify a liable autonomous vehicle for dispute settlement to prevent repudiation. \\
\end{itemize}

\section*{Blockchain Framework for Liability Attribution and Adjudication}
In this section, we discuss the details of the proposed permissioned BC based framework for liability attribution for autonomous vehicles. The most important feature of our framework is the partitioned communication to ensure data exchanges are based only on a need to know basis. \\
We utilize existing Public Key Infrastructure such as a certified authority (CA) to issue unique digital identities to entities to facilitate authenticated and authorized communication. However, given the privacy concern of an autonomous vehicle owner highlighted in the previous section, we propose a solution based on IEEE 1609.2 [10] whereby, an entity can communicate with other entities without revealing its identity. The IEEE 1609.2 security standard defines a vehicular public key infrastructure (VPKI) for secure vehicular communications to ensure that data exchanges are secured. Data integrity is protected by digital signatures along with a corresponding certificate sent with the data. The certificate is based on pseudonyms to preserve the privacy of the vehicle owner. According to a safety pilot model [11], the validity of a certificate is about 5 minutes thus the verification data sent by an autonomous vehicle would include the validity of the certificate. \\
Data exchanges between entities in our framework are stored in the BC and used for making liability decisions. They include: accident data collected by an autonomous vehicle to know if a product defect or driver error was responsible for the accident; instructions from an auto manufacturer to the autonomous vehicle to execute an update task to know if the negligence of the driver contributed to the accident; and the request for corroboratory evidence to facilitate payment of compensation. To ensure that data exchanges are relevant to intended entities and that entities are not privy to privacy related data, we use partitions in our framework. Partitions are communication segments that only allow intended entities to exchange relevant data contributing to evidence. For example,  keeping track of interactions between an autonomous vehicle and other interacting entities is envisaged to identify possible neglects of instruction by an autonomous vehicle owner. While such records for evidence might be vital to an insurance company to attribute liability or an auto manufacturer that also schedules maintenance for an autonomous vehicle, it is not always relevant to the dispute resolution authorities such as the government transport authority (\textit{GTA}) and legal authority (\textit{LA}). However for contested liability decisions, all evidence used in making initial liability decision is presented to the (\textit{GTA}) and legal authority (\textit{LA}) for final liability attribution. \\
We define two communication partitions in our framework: operational and decision partitions. Figure 2 shows the interaction of entities in both partitions. Entities in the operational partition, ($P_{1}$) include the insurance company, autonomous vehicle, service technician and auto manufacturer. Entities in the decision partition, ($P_{2}$) include the insurance company, auto manufacturer, government transport authority and legal authority. Entities in $P_{1}$ store data of relevant communications between interacting entities as contributing evidence in the BC to identify a potential case of negligence. They also store data generated from the autonomous vehicle when a collision event occurs. To attribute liability, data generated when a collision occurs is analyzed along with other data contributing to evidence stored in the BC such as instructions to execute an update from an auto manufacturer to an autonomous vehicle to identify a possible case of neglect. In $P_{2}$ contents of witness data are provided by road transport authority and legal authority as corroboratory evidence to the insurance company to facilitate liability decision making among liable entities in the operational partition, $P_{1}$  and to expedite claims processing. Also, liability decisions about collision events involving multiple vehicles and the identification of the liable vehicle is accomplished in $P_{2}$.  $P_{2}$ integrates all concerned  entities in the liability attribution process including auto manufacturers of vehicles involved in the accident and insurance companies of vehicle owners. In a multiple collision event such as a rear-ended collision, insurance companies and auto manufacturers send collision data received from their vehicles including a historical proof of their vehicle's past behavior stored in $P_{1}$ BC to the decision partition $P_{2}$ BC to expedite liability attribution. Historical proof is included to observe if the drive pattern of driver (human or machine) is consistent with the collision event to know if the accident was staged [12]. For example if historical behavior reveals that a driver is prone to hard brake slams, he could be blamed for the accident.\\ 
The roles of entities in our proposed framework are further classified as proposers and validators as shown in Table 1. Proposers are entities that submit data contributing to evidence while validators are entities that verify and validate transactions. They are responsible for distributedly maintaining the BC. Proposers in the operational partition $P_{1}$ include the service technician, auto manufacturer and autonomous vehicle. They store data contributing to evidence in the BC such as the maintenance report from a service technician, instruction to execute updates from an auto manufacturer and the collision data from an autonomous vehicle. Proposers in the decision partition $P_{2}$ include insurance companies of vehicle owners and auto manufacturers. They store evidence in $P_{2}$ BC  for liability attribution in the event of multiple collision or in the unlikely scenario when liability decisions in the operational partition are contested. Validators  in $P_{1}$ include the auto manufacturer, maintenance service provider and insurance company. Validators in $P_{2}$ include the road transport authority and legal authority. The motivation for the selection of validators vary in both partitions. In the operational partition validators are selected based on to the communication and storage capabilities of interacting entities. This is because a distributed ledger could grow  dramatically and this may cause significant communication and storage overhead for the resource constrained vehicle. In the decision partition, validators are selected based on the need to restrict access to sensitive privacy related data. \\
\begin{table}[ht]
\caption {Roles of Entities in the Proposed BC Framework} \label{tab:title}
\begin{tabular}{|| l | p{6cm} | p{6cm} ||}\hline
 \hline
 \multicolumn{3}{c} {Concerned Entities} \\ \hline
Role &  $P_{1}$ : Operational Partition & $P_{2}$ : Decision Partition \\
\hline
Proposer & Autonomous vehicle, Service technician, Auto manufacturer & Insurance company, Auto manufacturer  \\
\hline
Validator & Insurance company, Service technician, Auto manufacturer & Government transport authority,  Legal authority  \\
\hline 
\end{tabular}
\end{table}

\begin{figure*}[h]
\centering
\includegraphics[width=0.85\textwidth]{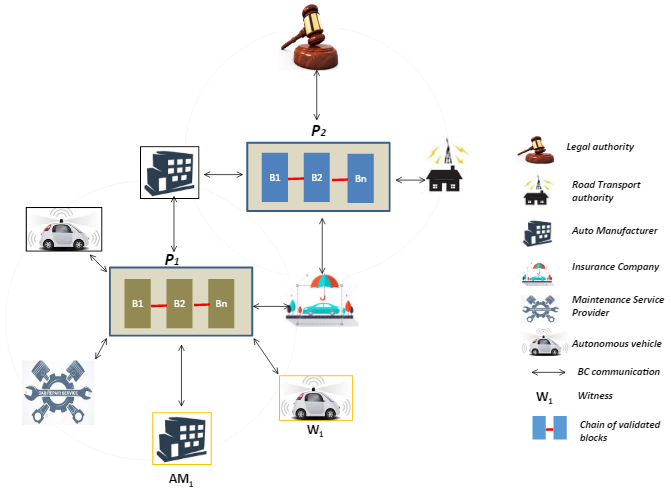}
\caption{Proposed Blockchain Based Framework.}
\end{figure*}

Figure 2 presents our proposed framework for liability attribution. In $P_{1}$ relevant interactions between a vehicle and its other interacting called transactions are sent to the BC by proposers, verified and committed to the current block \textit{cBlock} by validators. This includes the instructions to the vehicle to execute an update, go for a maintenance, the maintenance report from a service technician about a maintenance conducted on an autonomous vehicle and collision data generated by an autonomous vehicle in the event of an accident. Once received, validators verify the transaction by verifying the signature of the transaction proposer. Once successfully verified, the transaction is added to \textit{cBlock}. The process is similar in $P_{2}$. The difference however is the communication between interacting entities. In $P_{2}$, a request for evidence is sent to the BC by a proposer such as an insurance company. This request contains details of an accident generated by the vehicle involved in the accident. Validators verify the accident data, reveal the identity of vehicle owner and provide the accounts of witnesses to facilitate payment of compensation.\\
We next describe in details data contributing to evidence and further describe the process of verifying and validating transactions in our proposed framework. \\ \\ \\

\textbf{Transactions} \\
Transactions are communications between interacting entities and refer to data contributing to evidence in our framework. They are initiated by proposers and secured using cryptographic hash functions (SHA256), digital signatures and asymmetric encryption. In our framework, transactions could be signed by a single entity (called single signature transaction) or multiple entities called multiSig transactions. A MultiSig transaction ensures that parties to a transaction accept the transaction as valid. \\
The transactions considered in our proposed framework are: i) Event Safety Transaction which provides historical proof of the behavior of an autonomous vehicle; ii) Primary Evidence Transaction which contains actual collision data; iii) Notification Evidence Transaction which is used to keep track of negligence liability; and iv) Request Evidence which is used to obtain evidence to facilitate compensation payments.The structure of these transactions including data contained within a transaction is described below:
\begin{itemize}
\item \textbf{Event Safety Transaction \textit{(EST)}}: \textit{EST}  notifies validators of a likely collision. It is predicated on the event safety message generated by an autonomous vehicle when conditions in [13] are met such as hard braking of vehicle, detection of wrong way driving and detection of slippery road condition. EST is a single sign transaction by an AV and sent to $P_{2}$.  The structure of \textit{EST} is shown below.

\begingroup
    \fontsize{8pt}{12pt}
    \begin{align*}  
       EST = [ TID ||ts || ESM || TSdata || CertAV || Sig ]
    \end{align*}
\endgroup
\textit{TID} is the hash of \textit{EST}, ts is time of event occurrence, \textit{ESM} is the event safety message and contains forensic data such as location, speed, position. \textit{TSdata} is the last data in tamper proof storage before event. It contains the hash of video and picture files. The actual video and picture files are separately stored in a local storage and separately sent to $P_{1}$ validators. \textit{CertAV} is to ensure anonymous communication between an autonomous vehicle and other entities. \textit{CertAV} is different for every transaction generated by the vehicle and the validators verify its validity at transaction generation time. \textit{Sig} is the signature of the autonomous vehicle on the transaction. 

\item \textbf{Primary Evidence Transaction \textit{(PET)}}: \textit{PET} is a single sign transaction generated by autonomous vehicles in the event of an accident. It is transmitted by vehicles involved in the accident or witness vehicles. The structure of \textit{PET} is given below.
\begingroup
    \fontsize{8pt}{12pt}
    \begin{align*}  
        PET = [TID || edata|| TSdata || CertAV || Sig ]
    \end{align*}
\endgroup
\textit{edata}  is the evidence data generated by an autonomous vehicle when collision occurs and transmitted to $P_{1}$. The content of \textit{edata} is presented below: 
\begingroup
    \fontsize{8pt}{12pt}
    \begin{align*}  
        edata = [loc, TS, HVdata , TSdata , E(WVdata), h(edata)]
    \end{align*}
\endgroup
\textit{TS} is the event occurrence time, $HV_{data}$ is the recorded perception of the host vehicle and contains \textit{ESM} data. \textit{E($WV_{data}$)} provides  the encrypted testimonies of witness vehicles that were either involved in the collision or in camera and radio range of the collided vehicles. h(edata) is the hash of edata to assure its integrity.
\item \textbf{Notification Evidence \textit{(NE)}}: Keeping track of interactions between potential entities has been identified as important to identify scenarios of neglect on the part of a vehicle owner to carry out an instruction from an automaker or service technician. NE keeps track of instructions to the autonomous vehicle and execution of such instructions. We define three types of notification evidence transactions (update, maintenance and execute) to reflect specific interactions between entities. Given that vehicles have fast changing topologies and experience disconnection [14], instructions from validators are repeatedly sent until acknowledgement is received from vehicles.\\
\textbf{Update Transaction \textit{UT}}: \textit{UT} is an instruction from an auto manufacturer to an autonomous vehicle to execute an update. It is a multiSig transaction initiated by an auto manufacturer and completed by an autonomous vehicle to prevent repudiation. The structure of the transaction is given below:
\begingroup
    \fontsize{7pt}{10pt} 
    \begin{align*}  
        UT = [TID || h (UT)||PKAM ||SigAM ||SigAV ||CertAV ||Metadata]
    \end{align*}
\endgroup
\textit{h(UT)} is the hash of the update file and the metadata field contains details about the update and $PK_{AM}$ is the known public key of \textit{AM}. \\
\textbf{Execution Transaction \textit{ET}}: \textit{ET} is a follow up to \textit{UT}. The execution of \textit{UT} is a single sign transaction by an autonomous vehicle and linked to \textit{UT}. 
\begingroup
    \fontsize{8pt}{12pt}
    \begin{align*}  
        ET = [TID || P T I D  ||EXEC STAT || SigAV ||CertAV ]
    \end{align*}
\endgroup
\textbf{Maintenance Transaction \textit{MT}}: \textit{MT} describes the actions of a service technician on an autonomous vehicle. Maintenance of an autonomous vehicle occurs either when an auto manufacturer issues a maintenance request to the vehicle or when a vehicle owner needs to renew the registration of his vehicle.  The former is based on diagnostics reports periodically sent to the auto manufacturer by an autonomous vehicle. An auto manufacturer stores the hash of the request in $P_{1}$. In the latter, a maintenance service provider certifies the road worthiness of the vehicle and submits the report to the blockchain.\\
\item \textbf{Request Evidence Transaction \textit{(RET)}}: In a single collision scenario such as hit and run, \textit{RET} is a request from the validators in $P_{1}$ to $P_{2}$ validators for encrypted accounts of witnesses. It is also a request for identification of data owner for liability attribution purposes. In a multiple collision scenario, it is a single sign request for the identification of liable vehicle by proposers in $P_{2}$. 
\begingroup
    \fontsize{8pt}{12pt}
    \begin{align*}  
        RET = [TID ||edata || PKIC || SigIC ||CertAV]
    \end{align*}
\endgroup
\textit{edata} is the evidence data generated by the vehicle of the accident. \textit{PKIC} is the known public key of the insurance company. The auto manufacturing company  also submits its own request for evidence to obtain decrypted testimonies of witnesses for further forensic analysis. \\

\end{itemize}

Transactions sent to the blockchain are continuously verified and validated by validators until transactions in a block reach the maximum block size ($B_{Max}$). Once transactions reach ($B_{Max}$), the block is appended as the last block in the chain. The genesis block starts the BC and contains verification components (certificates) of the certificate issuing authority and is used by validators to verify transactions and by proposers to authenticate transactions.\\  \\ \\

\textbf{Transaction Verification} \\  
Once a transaction is sent to the BC, the process of verification begins. A transaction submitted by a proposer is verified by confirming that the transaction generator is authorized to communicate in that partition, Once this is completed, validators check if transaction conditions are met such as if a multiSigned transaction was signed by all concerned participants. For transactions generated by an autonomous vehicle, the validity of its signing certificate is also verified. 
In summary, we consider the following verification policies in our framework:
\begin{itemize}
\item \textbf{Completeness: } A transaction must satisfy this verification requirement to be logged to the blockchain. For example, a multiSig transaction must be signed by concerned entities to be considered complete and verifiable. 
\item \textbf{Authorization: } The proposer is authorised to transact in either the operational partition, $P_{1}$ or the decision partition $P_{2}$.
\item \textbf{Uniqueness: } A transaction can only be sent to the BC once by a transaction generator.
\end{itemize}
\textbf{Transaction Validation} \\
A block can contain $B_{Max}$ transactions. Transaction generation in our framework is however unpredictable because the  occurrence events such as accidents, maintaining a vehicle and sending software updates are not fixed. Given the unpredictable rate of transaction generation, it takes a significant amount of time for transactions to reach $B_{Max}$ and to validate the block. To address this, we propose a dynamic block validation process where successfully verified transactions are validated in the current block \textbf{cBlock}. \\
Upon a successful verification of a transaction, the transaction is validated in the current block \textit{(cBlock)} by computing the hash of the successfully validated transaction and the hash of \textit{(cBlock)}. By verifying computed hashes, validators reach consensus on \textit{(cBlock)}. This continues until \textit{(cBlock)} equals $B_{Max}$ at which point, the last validation of the last transaction in the block generates the permanent block ID and appended to the chain as the last block.  Once a transaction is successfully verified and validated, it becomes usable evidence for liability attribution. \\

\begin{table*}[ht]
\caption {Requirement for blockchain based architecture} \label{tab:title} 
\centering
\begin{tabular}{|p{3.0cm}|p{13cm}|}
\hline
Requirement & Methodology \\
\hline
Evidence integrity & \textit{TID} represents the hash of a transaction and used to ensure that a transaction originating from an entity is not altered. Also, we hash data contents in a transaction to prevent alteration. For example, \textit{edata} in the primary evidence transaction is hashed to identify an attempt to alter evidence (Section III). Through computation of hashes, validators reach consensus on current block state (\textit{cBlock}). \\
\hline
Secure storage & Successfully verified transactions are committed in the tamper resistant block and only transactions in blocks are used in making liability decisions. Also, every successfully verified transactions result in a change in the current block (\textit{cBlock}) and prevents alteration of evidence.  \\
\hline
Non-repudiation &  Transactions in \textit{cBlock} are signed by all transaction originators and verified by validators to prevent repudiation of actions and ensure auditability.\\
\hline
Decentralization & No single validator can independently make decisions on what data constitutes evidence to be used in attributing liability. Data contributing to evidence are collaboratively verified by validators in a partition and only successfully verified transactions can be used to make liability decisions. \\
\hline
Authorization & $P_1$ and $P_2$  restrict communication to only authorized entities in a partition and allow communication of transactions only on a need-to-know basis.  The genesis block in each partition contains the verification credentials of a Certification Authority (CA). The verification credential is unique to a partition thus allowing only members in a given partition to authenticate and verify transactions.  \\
\hline
Privacy & Transactions are signed using fresh signing keys and verified using the known public key of the CA. Also, given the enormous data generated by an autonomous vehicle, we preserve the privacy of vehicle owner via anonymous communication with other entities using security protocols proposed by IEEE 1609. 2 [6]. An autonomous vehicle is issued anonymous certificates each valid for only a predefined period. This offers conditional anonymity allowing a vehicles identity to stay hidden until it is required for dispute settlement. While verifying a transaction generated by an autonomous vehicle, validators also verify the validity of the certificate used for the transaction by checking if the timestamp on transaction falls within the certificate validity period.  \\
\hline
\end{tabular}
\end{table*}
\section*{Discussion}
In this section, we discuss the security of the architecture. We begin by evaluating how our framework satisfies the identified requirements and the resilience of our framework against identified attack capabilities of entities in our framework as shown in Table 2. \\
The security provided by the proposed architecture can be attributed to the use of blockchain technology. Transactions as well as data contained in transactions are secured using a hashing algorithm to ensure transaction and data integrity. Also, transactions and data contained within are secured using asymmetric encryption methods to restrict access to data in a transaction.\\
Given multiple potential sources of liability, likely liable entities become motivated to deny their actions when an accident occurs. Furthermore, an internal member of staff in the insurance company or  the road transport authority could be complicit in making an evidence unavailable. In the following, we demonstrate the resilience of the proposed blockchain architecture against such  attack capabilities. \\

\textbf{Evidence tampering: } Given that liability attribution in the auto insurance is not time critical, transaction generation is infrequent and bursty. As such, a current block \textit({cBlock}) does not get easily filled up. Therefore validation of blocks is delayed until the current block reaches maximum block capacity ($B_{Max}$) i.e. \textit{cBlock} = $B_{Max}$. The time before \textit{cBlock} = $B_{Max}$ offers a window of opportunity to a potential liable validator to tamper with \textit{cBlock} by deleting transactions thereby making it difficult for validators to reach consensus on \textit{cBlock} when equal to $B_{Max}$. Our proposed framework addresses this attack by validating transactions as soon as they are successfully verified. This results to a new $Block_{ID}$ after each validation process making it difficult for a validator to tamper with transactions in \textit{cBlock} without detection.  \\

\textbf{False information: } We describe two possibilities of false information and demonstrate how our framework is resilient against such attacks. Firstly, in the event of an accident such as a multiple collision scenario, an auto manufacturer could after receiving primary evidence collude with its autonomous vehicle by propagating misleading information to the decision partition ($P_{2}$) validators by altering the contents of host vehicle data in \textit{edata} and computing a new hash of \textit{edata} ($h_{new}$\textit{edata}) to evade liability. For the first scenario, the decision partition validators verify data sent by a rogue auto manufacturer against the data sent by an insurance company. This is done by computing the hash of \textit{edata}, from insurance company and auto manufacturer and checking the location and timestamp on both data. The outcome will reveal the malicious intention of the rogue auto manufacturer as the computed hash as well as the time stamp will differ. Secondly, in the event of a compromised software, an autonomous vehicle may produce authenticated messages that contain false information. We rely on the verification of data consistency. Consistency is achieved by comparatively verifying data with other witnesses involved in the accident i.e. correlating data of an auto manufacturer and insurance company of an involved vehicle with other involved vehicle's auto manufacturer and insurance companies in both time and space by checking timestamps and location on each data. \\

\textbf{Unavailable evidence: } An attacker such as a road transport authority (RTA) employee with vested interest in the outcome of a liability decision could be complicit in making  evidence unavailable for liability decisions. In our proposed framework, transactions sent to the blockchain are first verified and then validated by validators. Validation is a fundamental process in our framework which allows validators to reach consensus on the state of \textit{cBlock} and achieve block consistency. A successful validation of transaction always results in an updated state of the current block with a new $cBlock_{ID}$ making it difficult for the rogue RTA employee to make evidence unavailable.

\section*{Conclusion}
In this paper, we proposed a novel distributed digital forensics framework for the auto insurance liability model for the anticipated autonomous vehicles that provides untampered evidence for processing auto insurance claims and settling disputes. This framework eliminates the possibility of single point of trust allowing multiple participants to simultaneously agree on the evidence needed to process claims for compensation. The proposed framework is based on a permissioned blockchain that is partitioned to restrict communication to authorized participants. We presented a new liability attribution model for autonomous vehicle,  identified key requirements and described the threat model. We also described the structure of evidence including how it is verified and validated by entities in our proposed model. Furthermore, we conducted an analysis to show how our framework achieves identified requirements and how our proposal is resilient to possible attacks from concerned entities. Directions for future work includes an implementation of the architecture and evaluating its performance using potential real world collision scenarios. Our proposed architecture can be applied in other domains such as health insurance and accident reconstruction. 

\begin{singlespace}
\section*{References}\sloppy
\phantomsection
\label{csl:1}[1] N. Boudette, "{Autopilot cited in Death of Chinese Tesla Driver.}", Also available at: 
https://www.nytimes.com/2016/09/15/business/fatal-tesla-crash-in-china-involved-autopilot-government-tv-says.html

\phantomsection
\label{csl:2}[2]N. R Fulbright, "{Autonomous vehicles: The legal landscape of dedicated short range communication in the US, UK and Germany.}",  July, 2017.  

\phantomsection
\label{csl:3}[3] A. Greenberg, "{Hackers Remotely Kill a Jeep on the Highway- With me in it.}" Also available on: https://www.wired.com/2015/07/hackers-remotely-kill-jeep-highway/ 2015.

\phantomsection
\label{csl:4}[4] NRMA Insurance, "{Car Insurance Claims.}"  Also available on: https://www.nrma.com.au/claims/car-insurance

\phantomsection
\label{csl:5}[5] P. Jayachandran, "{Similarities Between Public and Private Blockchains}", Also available at: 
https://www.ibm.com/blogs/blockchain/2017/05/the-difference-between-public-and-private-blockchain/  May, 2017.

\phantomsection
\label{csl:6}[6] Hyperledger, "{Welcome to Hyperledger.}" Also available on: https://hyperledger-fabric.readthedocs.io/en/release/blockchain.html

\phantomsection
\label{csl:7}[7] R. Heijden, F. Engelmann, D. Modinger, F. Schonig and F. Kargl "{Blackchain: Scalability for Resource-Constrained Accountable Vehicle-to-X Communication.}" \textit{SERIAL’17: ScalablE and Resilient InfrAstructures for distributed Ledgers, December 11–15, 2017, Las Vegas, NV, USA. ACM, New York, NY, USA, 5 pages.}, 2017.

\phantomsection
\label{csl:8}[8] A. Dorri, M. Steger, S. Kanhere and R. Jurdak "{BlockChain: A Distributed Solution to Automotive Security and Privacy.}" \textit{IEEE Communication Magazine} 2017.

\phantomsection
\label{csl:9}[9] S. Rahman and U. Hengartner, "{Secure Crash Reporting in Vehicular Ad-hoc Networks.}" \textit{University of Waterloo}, 2017.

\phantomsection
\label{csl:10}[10] W. Whyte, A. Weimerskirch, V. Kumar and T. Hehn, "{A Security Credential Management System for 
 Communications.}" \textit{in Vehicular Networking Conference (VNC), 2013 IEEE.} IEEE 2013, pp. 1-8.

\phantomsection
\label{csl:11}[11] U. D. of Transportation, "{Safety Pilot Model Deployment. [online]}" Available on: safetypilot.umtri.umich.edu

\phantomsection
\label{csl:12}[12] DMV.org, "{How to handle staged car accidents}" Available on:https://www.dmv.org/insurance/how-to-handle-staged-car-accidents.php  \ 

\phantomsection
\label{csl:13}[13] ETSI, "{Intelligent Transport System (ITS), Vehicular Communications; Basic Set of Applications; Part 3: Specifications of Decentralized Environmental Notification Basic Service.}" \textit{ETSI TS}, 102 637-3 V1.1.1 (2010-09).

\phantomsection
\label{csl:14}[14] B.Paul, MD.Ibrahim, MD.Biskas, "{VANET Routing Protocols: Pros and Cons.}" \textit{International Journal of Computer Applications (0975 - 8887)}, Volume 20-No. 3, April 2011.
\end{singlespace}

\end{document}